\documentclass[aps,prc,twocolumn]{revtex4}
\usepackage{epsfig}
\usepackage{graphicx}
\newcommand{\be}{\begin{equation}}
\newcommand{\ee}{\end{equation}}
\newcommand{\bea}{\begin{eqnarray}}
\newcommand{\eea}{\end{eqnarray}}
\newcommand{\nn}{\nonumber\\ }
\newcommand{\muMS}{\bar\mu_{{\rm MS}}}
\begin{document}
\title{Hot QCD equations of state and relativistic heavy ion collisions
}
\author{Vinod Chandra}
\thanks{vinodc@iitk.ac.in}
\author{Ravindra Kumar}
\thanks{rojha@iitk.ac.in}
\affiliation{Department of Physics, IIT Kanpur, Kanpur-208 016, India.
}
 \author{V. Ravishankar}
\thanks{vravi@iitk.ac.in}
\affiliation{Department of Physics, IIT Kanpur, Kanpur-208 016, India
}
\affiliation{Raman Research Institute, Bangalore, 560080, India}
\begin{abstract}
We study two recently proposed  equations of state (EOS) which are obtained from high temperature QCD, and show how they  can be adapted to use them for making  predictions for relativistic heavy ion collisions. The method involves extracting  equilibrium distribution  functions for quarks and gluons from the EOS, which in turn will allow a determination of the transport and other bulk properties of the quark gluon plasma. Simultaneously, the method also yields a quasi particle description of interacting quarks and gluons. The first EOS is perturbative in the QCD coupling constant and has contributions  of $O(g^5)$. The second EOS is an improvement over the first, with contributions upto $ O(g^6 ln(\frac{1}{g}))$; it incorporates the nonperturbative hard thermal contributions. The interaction effects are shown to be captured entirely by  the effective chemical potentials for the gluons and the quarks, in both the cases. The chemical potential is seen to be highly sensitive to the EOS. As an application, we determine the  screening lengths which are, indeed the most important diagnostics for QGP. The screening lengths are  seen to behave drastically differently depending on the EOS considered., and yield, therefore, a way to distinguish the two equations of state in heavy ion collisions.
\end{abstract}


\maketitle
\vspace{2mm}
\section{Introduction}
Recent experimental results\cite{star,phn,pb,bhm} indicate that the 
quark gluon plasma has already been produced at RHIC, and that its behavior is not close to that of an ideal gas. Indeed, measurements of flow parameters
\cite{star}, and observations of jet quenching\cite{jet} have stimulated the theoretical interpretation that the QGP behaves like a nearly perfect fluid\cite{mjt}, characterized by a small value of the viscosity to entropy density ratio, lying in the range  $.1 \sim.3$. \cite{dtn,br,olli}; this range may be contrasted with the corresponding value for  liquid Helium (above superfluid transition temperature) which is close to ten \cite{dtson}. These observations signal the fact that the deconfined phase is strongly interacting, and are consistent with the lattice simulations\cite{lattice}, which predict a strongly interacting behavior even at temperatures which are a few $T_c$.  In an attempt to appreciate this surprising result, interesting analogies have been drawn with $ADS/CFT$ correspondence\cite{dtson} and also with  some strongly coupled classical systems\cite{shur1}.  In any case, the emergence of the strongly interacting behaviour puts into doubt the credibility of a large body of analyses which are based on ideal or nearly ideal behaviour of QGP.

 In this context, there is an interesting attempt by Arnold and Zhai \cite{arnold} and Zhai and Kastening\cite{zhai} who have determined the equation of state (EOS) of interacting quarks and gluons upto $ O(g^5)$ in the coupling constant . This strictly perturbative EOS, which we henceforth denote by EOS1, has been improved upon by Kajantie et al.\cite{kaj1,kaj2}  who have incorporated the contributions
from the  nonperturbative scales {\it viz.} $gT$ and $g^2 T$ and determined the EOS upto $O(g^6 ln(\frac{1}{g}))$\cite{ipp}. The latter will be denoted by EOS2.
  Subsequent studies \cite{blz1,blz2,rebh} have emphasized the relevance of the above EOS to study quark gluon plasma. One would naturally wish to compare these EOS with (the fully non-perturbative)
lattice results.  EOS2 has been found  \cite{kaj1} to be in   qualitative agreement with the lattice results.  It is not without interest to explore further whether  this qualitative agreement can be further quantified, and whether HTL improved EOS can describe the QGP produced in heavy ion collisions. It is worthwhile noting the earlier attempts  \cite{rebhan1,rebhan2,JPP} that have been made to determine thermodynamic quantities such as entropy and the specific heat $c_v$ in improved perturbative approaches to QGP.

    On the other hand, it is by now well established that the semiclassical approach is a convenient way  to study the bulk properties of QGP\cite{vr1,vr2,vr3,CM1,CM2}, since they automatically incorporate  the HTL effects\cite{CM1,CM2}. There is a wealth of results which have been obtained within this framework\cite{vr1,vr2,vr3}, where the nonperturbative
    features manifest as effective mean color fields. These color fields  have the dual role of producing
    the soft and semisoft partons, apart from modulating their interactions. The emergence of
    such effective field degrees of freedom, together with a classical transport has been indicated earlier by Blaizot and Iancu \cite{blaizot}.
 
In this context, it is pertinent to ask if one could 
use heavy ion collisions to distinguish the various EOS and pick up the right one, by employing the semiclassical framework involving an appropriate kinetic equation.The purpose of this paper is to explore such possibilities. As a first step in this direction, we shall show how the distribution functions underlying the  proposed EOS can be extracted with a minimal ansatz, {\it viz}, effective chemical potentials for quarks and gluons. Once the distribution function is obtained, it can be used to study the bulk properties of the system such as chromo responses including the ubiquitous Debye Mass. Postponing all the other applications to a future work, we shall concentrate on determining the Debye mass through this procedure. 
 As mentioned, we focus on EOS1 and EOS2. 
 Both of them have been proposed for the case when the baryon number density vanishes. The corresponding chemical potentials are hence set to zero. There exist generalizations of the above EOS,
proposed by Vuorinen \cite{prd68} and more recently by Ipp et al \cite{ipp}, which allow for a finite
baryon number.  The  two sets are applicable to distinct physical situations; the former (EOS1 and EOS2)
are relevant to the QGP in the central midrapidity region of URHIC while the
works of Ref. \cite{prd68,ipp} are applicable to peripheral collisions and /or  when the so called nuclear transparency is only partial. An application of the above EOS to URHIC will be taken up separately.

  This paper is organized as follows. In the next section, we  extract the distribution functions for the gluons and the quarks from EOS1 and EOS2.  We consider the pure gluonic case separately from the full QCD, by first setting $N_f=0$. The (interacting) quark sector is then dealt with. In section III, the Debye mass is determined by employing the semiclassical method developed by Kelly et.al \cite{CM1,CM2}. In section III(B), we compare our results on screening length for EOS1 and EOS2 with the recent lattice results. We summarize the results and conclude in section IV. The appendix contains some details of calculations which are not explicitly given in the main text; it also lists some useful integrals.

 \section{Extraction of the distribution functions}

Recently Arnold et. al \cite{arnold} have derived an equation of state (EOS1)  for high temperature QCD up to $O(g^5)$.  EOS1 reads,
\begin{widetext}
\bea
\label{eqn1}
P^{(1)}&=&\frac{8\pi^2}{45\beta^4}\bigg \lbrace (1+\frac{21N_f}{32})-\frac{15}{4}(1+\frac{5N_f}{12})\frac{\alpha_s}{\pi}
+30(1+\frac{N_f}{6})(\frac{\alpha_s}{\pi})^{\frac{3}{2}} \nonumber\\
&&
+\bigg[(237.2+15.97N_f-0.413 N_f^2 +\frac{135}{2}(1+\frac{N_f}{6})\ln(\frac{\alpha_s}{\pi}(1+\frac{N_f}{6}))\nonumber\\
&&
-\frac{165}{8}(1+\frac{5N_f}{12})(1-\frac{2N_f}{33})\ln[\frac{\muMS\beta}{2\pi}]\bigg](\frac{\alpha_s}{\pi})^2\nonumber\\
&&+(1+\frac{N_f}{6})^{\frac{1}{2}}\bigg[-799.2-21.99N_f-1.926N_f^2\nonumber\\
&&+\frac{495}{2}(1+\frac{N_f}{6})(1+\frac{2N_f}{33})\ln[\frac{\muMS\beta}{2\pi}]\bigg](\frac{\alpha_s}{\pi})^{\frac{5}{2}} \bigg \rbrace
+O(\alpha_s)^{3}\ln (\alpha_s)). \nonumber\\
\eea
\end{widetext}
EOS1 has been subsequently improved by Kajantie et.al. \cite{kaj1,kaj2} who proposed another equation of state (EOS2) by  improving the accuracy to the next order in the coupling constant, and also included the HTL effects; recall that the latter are essentially nonperturbative, 
and contain contributions from scales $T,~gT,~{\rm and}~ g^2T$.  EOS2, which is thus determined
upto $O(g^6 ln\frac{1}{g})$, has the form
\begin{widetext}
\bea
\label{eqn2}
P^{(2)}&=&P^{(1)}+\frac{8\pi^2}{45}T^4 \biggl[1134.8+65.89 N_f+7.653 N_f^2\nn
&&-\frac{1485}{2}\left(1+\frac{1}{6} N_f\right)\left(1-\frac{2}{33}N_f\right)
\ln(\frac{\muMS}{2\pi T})\biggr]\left(\frac{\alpha_s}{\pi}\right)^{3}
\ln \frac{1}{\alpha_s}\,.
\eea
\end{widetext}
In the above expressions, $N_f$ is the number of fermions, $\alpha_s=g^2/(4\pi)$ is the strong coupling constant and $\muMS$ is the renormalization scale parameter in the $\overline {{\rm MS}}$ scheme. Note that $\alpha_s$ runs with $\beta$ and $\muMS$. As remarked, the
utility of this EOS in the context of QGP thermodynamics has been discussed earlier by Rebhan\cite{rebh}.

We now set to determine equilibrium distribution functions $<n_{g,f}>$  for the gluons and the quarks such that they would yield the EOS given above. The ansatz for the determination involves retaining the ideal distribution forms, with the chemical potentials $\mu_{g}$ and $\mu_f$ being  free parameters. Note that for the massless quarks ( $u$ and $d$) which we consider to constitute
the bulk of the plasma, $\mu \equiv 0$ if they were not interacting. This approach is of course not novel, since it underlies many of the ideas that attempt to describe the interaction effects in terms of the quasiparticle degrees of freedom. In the present context,
we refer the reader to Ref. \cite{pesh,allton}, where an attempt is made to describe the lattice results in terms of  effective mass for the partons.

We pause to note that the chemical potentials which we introduce are not the same as those which yield a nonzero baryon number density, as e.g., in \cite{prd68,ipp}. Here, the chemical poetntial merely serves to map the interacting quarks and gluons at zero baryon number chemical potential to noninteracting quasiparicles, {\it viz.}, the dressed quarks and gluons. Their interpretation is, therefore, more akin to
the effective mass, albeit as functions of the renormalization scale and temperature as we show below.
Thus, the baryon number density of the plasma continues to vanish.

As the first step in our approach, we express  the   EOS in the  form
\be
\label{eqn2.1}
  P=P^I_g +P^I_q + \Delta P_g + \Delta P_{f}
\ee
\\
The first two terms in the RHS of Eq.(\ref{eqn2.1})  are identified with the distributions of an ideal gas of quarks and gluons. The effects of the interaction in pure QCD  are represented by $\Delta P_g$ and the residual interaction effects, by $\Delta P_{f}$. 
For the EOS which we are interested in, the identification of the above terms is straight forward. $\Delta P_g$ can be identified by first setting $N_f=0$ and then subtracting the ideal part. The residual term is naturally identified as $\Delta P_{f}$ after subtracting the ideal part for quarks.
In general the form of EOS (see Eq.(\ref{eqn1}) and Eq.(\ref{eqn2}))
\bea
P&=&\frac{8\pi^2}{45}~\beta^{-4}~[ 1 + A(\alpha_s) + B(\alpha_s)\ln\frac{\muMS\beta}{2\pi}+ \frac{21}{32}N_f \nn &&+ C(\alpha_s,N_f)
+ D(\alpha_s,N_f)\ln \frac{\muMS \beta}{2\pi}]
\eea
where
\bea
\label{eqn3}
 P^I_g &=&\frac{8\pi^2}{45}\beta^{-4} \nn
 P^I_f &=&\frac{8\pi^2}{45}\beta^{-4} \frac{21}{32}N_f \nn
 \Delta P_g &=& A(\alpha_s) + B(\alpha_s)\ln\frac{\muMS\beta}{2\pi} \nn
 \Delta P_f &=& C(\alpha_s,N_f) + D(\alpha_s,N_f)\ln \frac{\muMS \beta}{2\pi}
\eea

For EOS1 the coefficients $A, B, C, D$ are denoted with a prime and are given by
\begin{widetext}
\bea
\label{eqn4}
A'(\alpha_s(N_f))&=&-\frac{15}{4}\frac{\alpha_s}{\pi}+30(\frac{\alpha_s}{\pi})^{\frac{3}{2}}
 +(237.2 +\frac{135}{2}\log(\frac{\alpha_s}{\pi})(\frac{\alpha_s}{\pi})^2 -799.2(\frac{\alpha_s}{\pi})^{\frac{5}{2}}
\nn
B'(\alpha_s(N_f))&=&-\frac{165}{8}(\frac{\alpha_s}{\pi})^2 +\frac{495}{2}(\frac{\alpha_s}{\pi})^{\frac{5}{2}}\nn
C'(\alpha_s(N_f),N_f)&=&-\frac{15}{4}(1+\frac{5}{12}N_f)\frac{\alpha_s}{\pi}    +30((1+\frac{1}{6}N_f)(\frac{\alpha_s}{\pi})^{\frac{3}{2}}+[237.2+\nn
&&
15.97N_f-0.413N_f^2+\frac{135}{2}((1+\frac{1}{6}N_f)\ln[\frac{\alpha_s}{\pi}(1+\frac{1}{6}N_f)](\frac{\alpha_s}{\pi})^2+\nn
&&
(1+\frac{1}{6}N_f)^(1/2)[-799.2 -21.99N_f-1.926N_f^2](\frac{\alpha_s}{\pi})^{\frac{5}{2}}-A'(\alpha_s(N_f))
\nn
D'(\alpha_s(N_f),N_f)&=&-\frac{165}{8}(1+\frac{5}{12}N_f)(1-\frac{2}{33}N_f)(\frac{\alpha_s}{\pi})^2\nn &&+
\frac{495}{2}(1+
\frac{1}{6}N_f)(1-\frac{2}{33})(\frac{\alpha_s}{\pi})^{\frac{5}{2}}-B'(\alpha_s(N_f))
\eea
whereas for EOS2 the coefficients can be written in terms of the above primed coefficients for EOS1, as
\bea
\label{eqn6}
A(\alpha_s(N_f)) &=& A'(\alpha_s(N_f)) +1134.8(\frac{\alpha_s}{\pi})^3\log(\frac{1}{\alpha_s})
\nn
B(\alpha_s(N_f)) &=& B'(\alpha_s(N_f)) -\frac{1485}{2}(\frac{\alpha_s}{\pi})^3\log(\frac{1}{\alpha_s})
\nn
C(\alpha_s(N_f),N_f) &=& C'(\alpha_s(N_f),N_f) +(65.89N_f + 7.653N_f^2)(\frac{\alpha_s}{\pi})^3\log(\frac{1}{\alpha_s})
\nn
D(\alpha_s(N_f),N_f) &=&D'(\alpha_s(N_f),N_f)-\frac{1485}{2}[(1+\frac{1}{6}N_f)(1-\frac{2}{33}N_f)-1](\frac{\alpha_s}{\pi})^3\log(\frac{1}{\alpha_s}). \nn
\eea
\end{widetext}
We seek to parametrize the contributions from all the non-ideal coefficients in terms  the chemical potentials $\mu_g$ and $\mu_f$ for gluons and quarks respectively. Since the EOS have been proposed at high $T$, with their validity being
at temperatures greater than $2T_c$ \cite{shaung},  we treat the dimensionless quantity $\tilde\mu_{g,f}  \equiv \beta \mu_{g,f}$ perturbatively. This approximation needs to  be implemented self consistently, and 
accordingly, we expand the  the grand canonical partition functions for gluons and quarks as a  Taylor series in $\tilde\mu_{g,f} $. We obtain the following expressions:
\bea
\label{eqn7}
\log( Z_ g) &=& \sum_{k=0}^\infty (\tilde\mu_g)^k \partial^{k}_{\tilde\mu_g} \log(Z_g)|_(\tilde\mu_g=0) \nn
\log(Z_q) &=& \sum_{k=0}^\infty (\tilde\mu_f)^k \partial^{k}_{\tilde\mu_f} \log(Z_f)|_{(\tilde\mu_f=0)}
\eea
\\
where, $ Z_g $ and $ Z_f $  are given by
\bea
\label{eqn9}
Z_g &=&\prod_p \frac{1}{(1-\exp(-\beta\epsilon_p+\tilde\mu_g))},
\nn
 Z_q &=&\prod_p \frac{1}{(1+\exp(-\beta\epsilon_p+\tilde\mu_f))}.
\eea
We determine $Z_g$ and $Z_q$ (defined in Eq.(\ref{eqn7})) up to $O(\tilde\mu_{g,f})^3$. 
The truncation is seen to yield an accuracy of $\sim$ ten percent when we consider EOS1. However, for the more physical EOS2, the agreement is within one percent.
The gluon chemical potential $\mu_g$ gets determined by

\be
\label{eqn10}
\frac{A^{(3)}_g}{3!}(\tilde\mu_g)^3 + \frac{A^{(2)}_g}{2!}(\tilde\mu_g)^2 +A^{(1)}_g\tilde \mu_g-\Delta P_g\beta V = 0.
\ee
\\
Similarly, the equation determining $\mu_f$ reads:
\\
\be
\label{eqn11}
\frac{A^{(3)}_f}{3!} (\tilde\mu_f)^3 + \frac{A^{(2)}_f}{2!}(\tilde\mu_f)^2 +A^{(1)}_f\tilde \mu_f-\Delta P_f\beta V = 0
\ee

The coefficients $A^{(n)}_g$ and $A^{(n)}_f$ are given by
\bea
\label{eqn12}
A^{(n)}_g=\partial^n_{\tilde\mu_g}\log(Z_g)|_{(\tilde\mu_g = 0)} \nn
A^{(n)}_f=\partial^n_{\tilde\mu_f}\log(Z_f)|_{(\tilde\mu_f = 0)}
\eea
The explicit forms of $A^n$ up to $n=3$ are listed in the Appendix.

\subsection{Explicit evaluation of the chemical potential}
Before we discuss the solution of Eq.(\ref{eqn10}) and Eq.(\ref{eqn11}), it is instructive to evaluate $\mu_{g,f}$ with just the linear and quadratic terms for the sake of comparison.
 In the linear order, the solutions read
\bea
\label{eqn13}
\tilde\mu_g &=& \frac{8\pi^2}{45(A^{(1)}_g)}[A(\alpha_s) + B(\alpha_s)\ln\frac{\muMS\beta}{2\pi}] \\
\tilde\mu_f &=& \frac{8\pi^4}{45(A^{(1)}_f)}[C(\alpha_s)+D(\alpha_s)\ln\frac{\muMS\beta}{2\pi}]
\eea
while, in the next to the leading order, the solution has the form
\be
\label{eqn14}
\tilde\mu_g = -\frac{A^{(1)}_g}{A^{(2)}_g}\pm \sqrt((\frac{A^{(1)}_g}{A^{(2)}_g})^2+C_g)
\ee

\be
\label{eqn15}
\tilde \mu_f = -\frac{A^{(1)}_f}{A^{(2)}_f}\pm \sqrt((\frac{A^{(1)}_f}{A^{(2)}_f})^2+C_f),
\ee
where, $C_g={2\Delta P_g\beta V}/{A^{(2)}_g}$ and
       $C_f={2\Delta P_f\beta V}/{A^{(2)}_f}$. Finally, the exact solutions  can be   obtained by using well known algebraic techniques. Since the explicit algebraic solutions do not have an illuminating form, we show the solutions graphically instead
in the next subsection.

The distribution functions for the gluons and the quarks get determined, in terms of the chemical potentials, through 
\bea
\label{eqn16}
\langle n_ g\rangle_p= \frac{\exp(-\beta \epsilon_p+\tilde\mu_g)}{1-\exp(-\beta\epsilon_p+\tilde\mu_g)}
\nn
\langle n_ f\rangle_p = \frac{\exp(-\beta\epsilon_p+\tilde\mu_f)}{1+\exp(-\beta\epsilon_p+\tilde\mu_f)}
\eea

The extraction of the distribution functions is, nevertheless, incomplete. For, the EOS  --and hence the chemical potentials -- depend on the renormalization scale. On the other hand, the physical observables should be scale independent. We circumvent the problem
by trading off the dependence on $\muMS$ to a dependence on the critical temperature $T_c$.
To that end, we exploit  the temperature dependence  of  the coupling constant  $\alpha_s(T)$ \cite{shaung,comment1} and of renormalization scale:
 
\bea
\label{eqn17}
\muMS(T)&=& 4\pi T \exp(-(\gamma_E+1/22)\nn
\alpha_s (T) &=& \frac{1}{8\pi b_{0}\log(T/\lambda_T)} = \alpha_s (\mu^2) \vert_{\mu={\muMS}(T)} 
\nn
\lambda_T &=& \frac{\exp(\gamma_E+1/22)}{4\pi}\lambda_{MS}
\nn
\eea
where,
$b_0 ={33-2N_f}/{12\pi}$ and $\lambda_{MS}=1.14 T_c$. With this step, the distribution functions get determined completely, and are obtained as functions of $T/T_c$.  

We note that the results presented below, being valid for $T > 2T_c$, need to be supplemented by
a similar analysis for EOS which are valid for $T \sim T_c$. Such an analysis does indeed exist, along the lines of this paper \cite{allton}, who have considered the lattice EOS.
They do not determine the Debye mass, but focus on the impact of the EOS on the flow parameters
in heavy ion collisions.

Although, EOS1 and EOS2 have been computed within the framework of weak coupling technique  but they give convergent results 
for the temperature ranges which are more than $5 T_c$. We shall see in the next Subsection that these equations of state 
are far away from their ideal behaviour up to the extent that they can be utilized to make definite predictions for QGP.

\subsection{Hot QCD EOS vs Ideal EOS}
As a warm up, we compare EOS1 and EOS2  with the ideal EOS by plotting the ratio
$
R \equiv \frac{P}{P_q^I +P_g^I},$  as  functions of temperature, in Figs.1,2.

\begin{figure}[hbt]
\begin{center}
\includegraphics[scale=.7]{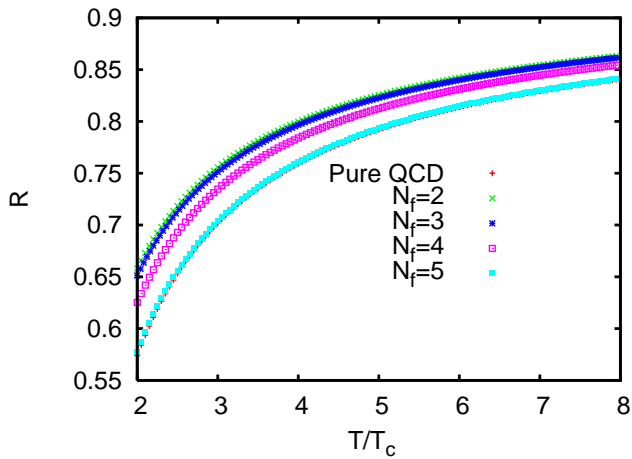}
\caption{(color online) Behaviour of R with temperature for EOS1}
\end{center}
\end{figure}

The most striking feature that we see  is the large sensitivity to the inclusion of the $g^6 ln(\frac{1}{g})$ contributions. It is most pronounced in the behaviour of pure QCD where major qualitative and quantitative differences appear:
(i) For EOS1, $R$ increases with $T$, in contrast to EOS2 which where it decreases from above,
approaching the same asymptotic value  for large $T$; (ii) interestingly,  the nonperturbative (and higher order) corrections  makes the system less non-ideal. Indeed, EOS1 yields values of $R$ which are  $10-45$ percent away from the ideal value 1, in contrast to EOS2 (see Fig2), for which  $R$ is only  $2-8$ percent away from the ideal value.
\begin{figure}[hbt]
\begin{center}
\includegraphics[scale=.7]{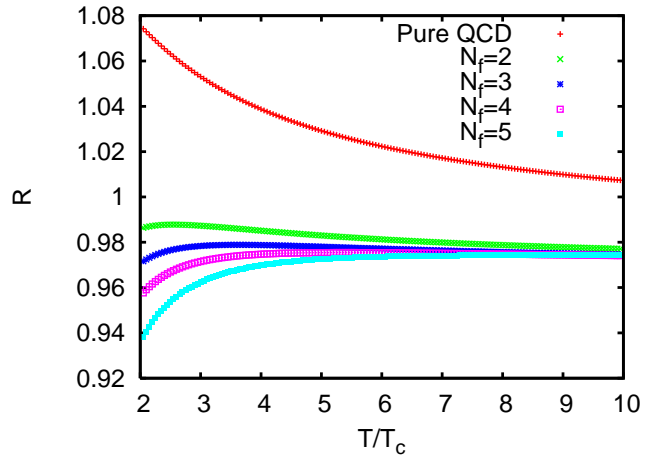}
\caption{(color online) Behaviour of R with temperature for EOS2}
\end{center}
\end{figure}
Incidentally, the above observation implies that the expansion in Eq.(\ref{eqn7}) works better
for EOS2 than the former.   This will be reflected later in the behavior of effective chemical potentials with temperature. We shall further see that, smaller the value of $\vert\tilde\mu_{g,f}\vert$, better will be the approximation.

\subsection{ The chemical potentials}

The variation of the effective chemical potentails with renormalization scale at a fixed temperature has already been studied in
 reference \cite{zhai}. We prefer to recast it into a dependence of $\tilde\mu_{g,f}$  on  $\frac{T}{T_c}$ (see the previous subsection) since it is more relevant to the study of QGP in heavy ion collisions.
This is shown in  Figs. $3- 8$, where the contributions coming from linear, quadratic and cubic approximations in the Taylor series(Eq.(\ref{eqn7})) are individually displayed, for both EOS1 and EOS2. These figures exhibit, in essence, all the interaction effects.
\begin{figure}[hbt]
\begin{center}
\includegraphics[scale=.7]{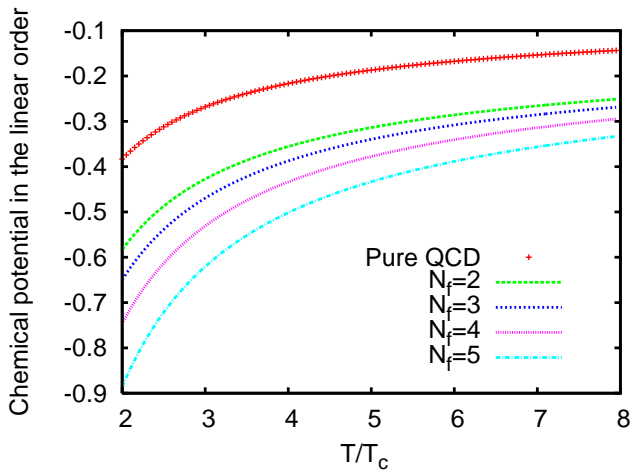}
\caption{(color online) Effective chemical potentials at the linear order for EOS1}
\end{center}
\end{figure}
\begin{figure}[hbt]
\begin{center}
\includegraphics[scale=.7]{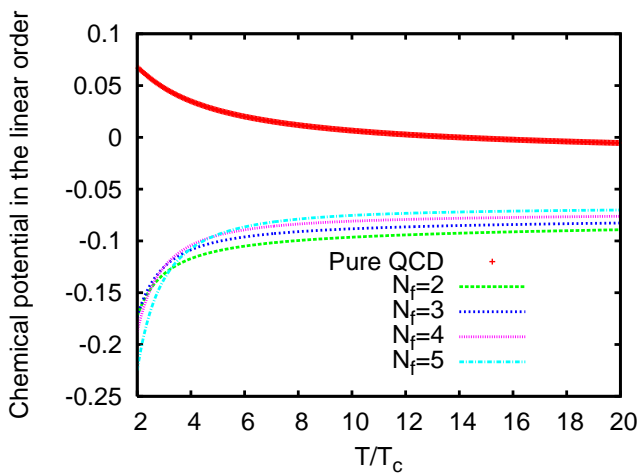}
\caption{(color online) Effective chemical potentials at the linear order for EOS2}
\end{center}
\end{figure}

A number of features emerge from examining Figs. $3-8$. Consider EOS1 first. Here,
the linear approximation  does reasonably well for pure QCD, but fails badly in the quark sector. The chemical potential is negative in both the sectors, and approaches the ideal value
asymptotically from below. In contrast, EOS2 leads to a  different behaviour: $\tilde\mu_g$ starts
with a small positive value at $T \sim 2T_c$, and stays essentially so until $T\sim 13T_c$,  and switches sign to acquire a small negative value. Since the magnitude remains less than
0.1 throughout, the deviation from the ideal behaviour is minimal. $\tilde\mu_f$ remains negative (with the maximum magnitude  $\sim$ 0.25 at $T=2T_c$),
 \begin{figure}[hbt]
\begin{center}
 \includegraphics[scale=.7]{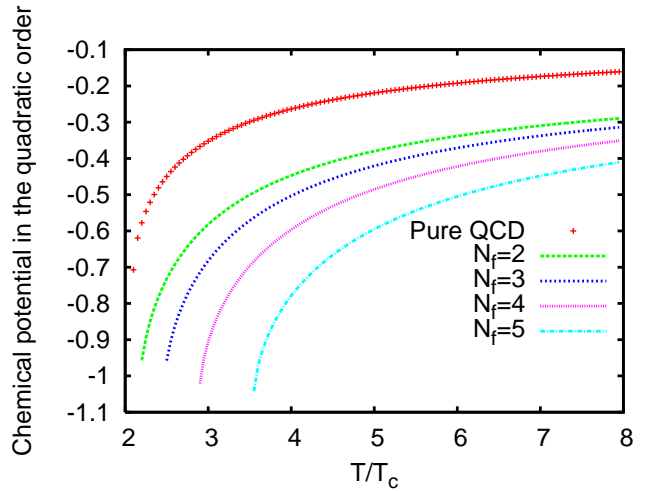}
\caption{(color online) Effective chemical potentials at the quadratic order for EOS1}
\end{center}
\end{figure}
\begin{figure}[hbt]
\begin{center}
\includegraphics[scale=.7]{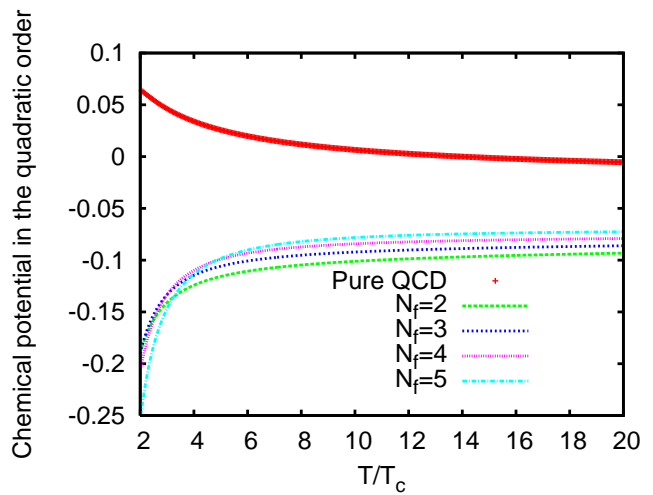}
\caption{(color online) Effective chemical potentials at the quadratic order for EOS2}
\end{center}
\end{figure}
which is about a factor four smaller in comparison with the corresponding value from EOS1.
The interaction effects get manifestly stronger as we increase the number of flavors.
\begin{figure}[hbt]
\begin{center}
\includegraphics[scale=.7]{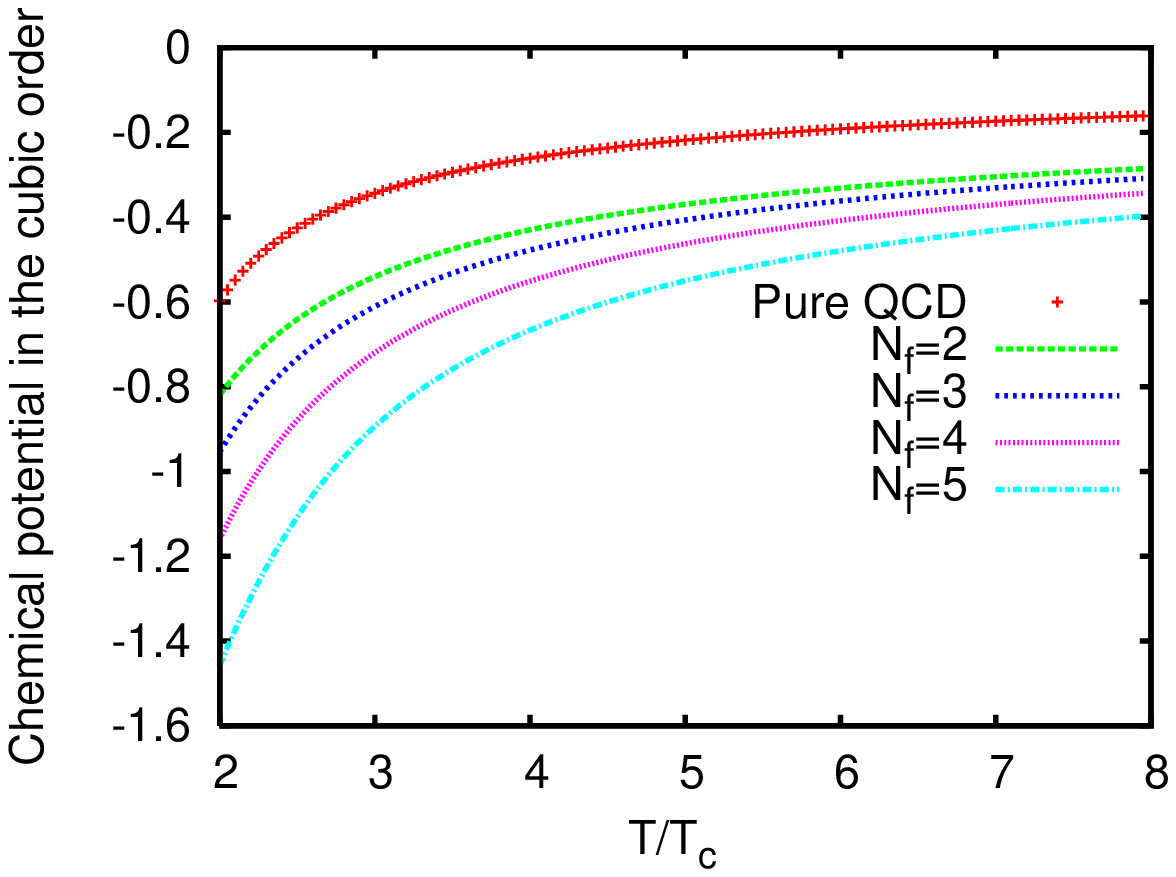}
\caption{(color online) Effective chemical potentials at the cubic order for EOS1}
\end{center}
\end{figure}
\begin{figure}[hbt]
\begin{center}
\includegraphics[scale=.7]{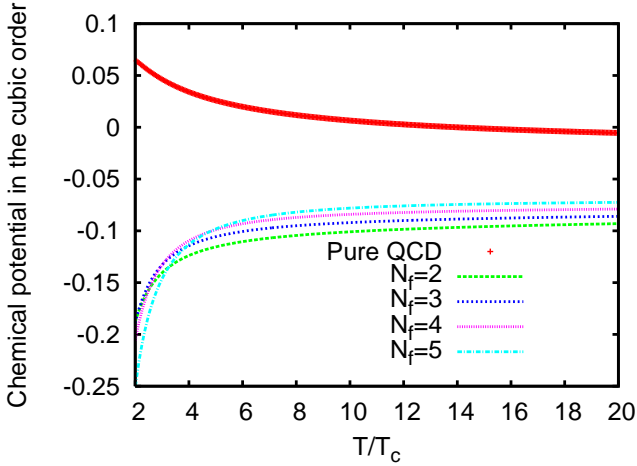}
\caption{(color online) Effective chemical potentials at the cubic order for EOS2}
\end{center}
\end{figure}

It is significant that the ideal value is not reached even at $T \sim 10T_c$, which indicates that the phase remains interacting. We also note that our method of extracting the chemical potential works more efficiently for EOS2, as indicated by small corrections from higher order terms to the linear estimate.

\section{The Debye mass and screening length}
The extraction  of the equilibrium distribution functions affords a determination of the
Debye mass, via the semiclassical transport theory \cite{CM2}. The Debye mass controls
the number of bound states in heavy $q\bar{q}$ systems, yields the extent of 
 $J/\Psi$ suppression in heavy ion collisions, provided that we have a reliable
estimate of the temperature of the plasma. Even otherwise, the qualitative significance of the Debye mass cannot be over estimated since  the deconfined phase remains strongly interacting
even at large $T$.

The determination of $m_D$ is straightforward if we employ the classical transport theory \cite{CM2}. It is simply given by

\be
\label{eqn18}
M^2_{g,f} =  g^{\prime 2} C_{g,f} \int \frac{d}{dp_0}<n_{g,f}> d^3p.
\ee
\\
The above expression has to be used cautiously, though. The coupling constant $g^{\prime}$ in eq.\ref{eqn18}
has a phenomenological character, and should not be confused with the fundamental constant
$g$ appearing in the EOS. Keeping this in mind, we recall that if the plasma were to be comprised of ideal massless
partons, the Debye mass would be given by
\\
\be
\label{eqn19}
M^2_{id}=M^2_{g,id}+M^2_{f,id} \equiv \frac{(N+N_f/2)}{3} g^{\prime 2}\beta^{-2}.
\ee
The hot  QCD EOS modify the above expression; It is easy to see, from Eqs.(\ref{eqn18}, \ref{eqn19}) that the new Debye masses, scaled with respect to their respective ideal values get determined, in terms of the standard
 PolyLog functions\cite{comment2} by
\\
\bea
\label{eqn20}
\frac{M^{2}_{g,hot}}{M^2_{g,id}}= \frac{6}{\pi^2} PolyLog[2,\exp(\tilde\mu_g)]\equiv F_1(\tilde\mu_g)
\nn
\frac{M^{2}_{f,hot}}{M^2_{f,id}}=-\frac{12}{\pi^2} PolyLog[2,-\exp(\tilde\mu_f)]\equiv
F_2(\tilde\mu_f).
\eea
\\
Consequently, the expression for the total relative mass is obtained as
\be
\label{eqn21}
\frac{M^2_{hot}}{M^2_{id}}=\frac{(\frac{N}{3} F_1(\tilde\mu_g)+\frac{N_f}{6} F_2(\tilde\mu_f))}{(N/3 + N_f/6)}.
\ee
It is, however,  more convenient to plot the inverse debye mass i.e, the screening length as a function of $T/T_c$.

\subsection{Relative screening lengths}
We first establish the notations.
Let $\lambda_h$ denote the screening length generated by the  hot EOS. Let $\lambda_{id}$ be the
screening length of an ideal qgp. It is convenient to consider also  the
contribution coming from the pure QCD sector, whose screening lengths we denote by
$\lambda_h^g$ and $\lambda_{id}^g$ respectively.

\begin{figure}[hbt]
\begin{center}
\includegraphics[scale=.7]{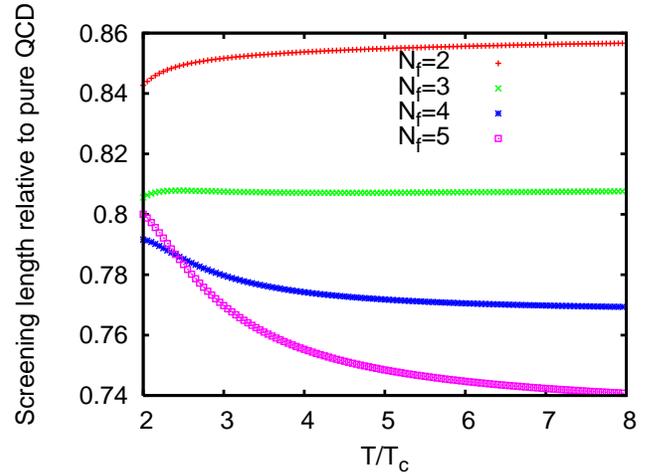}
\caption{(color online) The relative Debye Screening Length ${\cal R}_{h/g}$ for EOS1 as a function of temperature. Note that it is $\ge$ 0.7.}
\end{center}
\end{figure}

The behaviour of the screening lengths is shown in Figs.9 -12. As in the case of the chemical
potentials, the dependence on the order of perturbation is striking here as well. For EOS1
where the contributions upto $O(g^5)$ are included, the screening lengths in the full QCD as well as pure QCD remain nonzero. The dominant contribution is from the gluonic sector, which dominates over the quark sector, as may be seen in Fig.9 where we plot the ratio
${\cal R}_{h/g} ={\lambda_h}/{\lambda_h^g}$, which is in excess of 0.7 throughout. Note, however, that
the relative dominance gets weaker as we increase the number of flavours. Fig.10 shows the variation of the ratio ${\cal R}_{h/id}={\lambda_h}/{\lambda_{id}}$ as a function of temperature. Interestingly, the interaction is seen to weaken the screening, and so does an increase in the number of flavors.

\begin{figure}[hbt]
\begin{center}
\includegraphics[scale=.7]{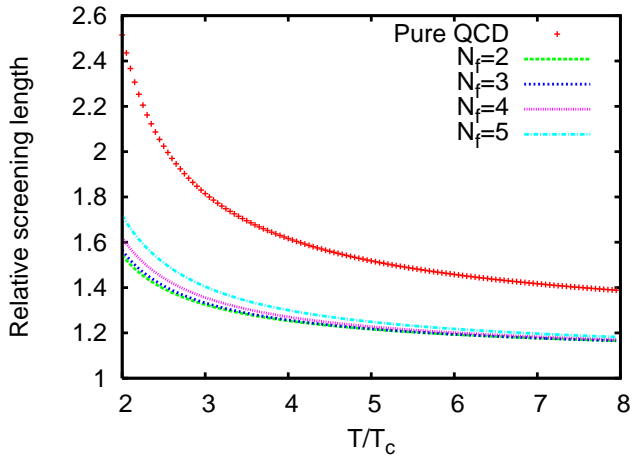}
\caption{(color online) The relative screening length${\cal R}_{h/id}$ for  EOS1 as a function of temperature.}
\end{center}
\end{figure}

\begin{figure}[hbt]
\begin{center}
\includegraphics[scale=.7]{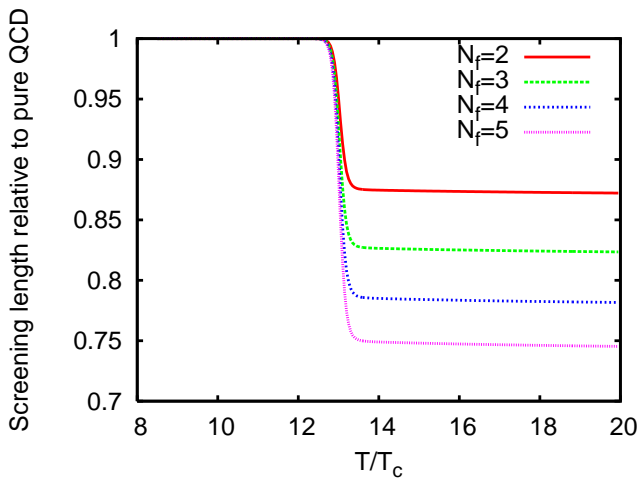}
\caption{(color online) The relative Debye Screening Length ${\cal R}_{h/g}$ for EOS2 as a function of temperature. Note that it stays at 1 all the way upto 13$T_c$.}
\end{center}
\end{figure}

These results are in sharp contrast with the case of
EOS2, which we recall has nonperturbative $O(g^6 ln(\frac{1}{g}))$ contributions. These are shown
in Figs.11 and 12. It is clear from Fig 11 that the contribution from the pure gluonic sector saturates the contribution to the screening all the way upto temperatures $T \sim 13T_c$, and drops sharply thereof. This
feature is reinforced by Fig.12 where the ratio ${\cal R}_{h/id}$ stays at zero
between $2T_c$ and $12-13T_c$. It is of a purely academic interest that the screening length
should become non-zero beyond $12T_c$.

\begin{figure}[hbt]
\begin{center}
\includegraphics[scale=.7]{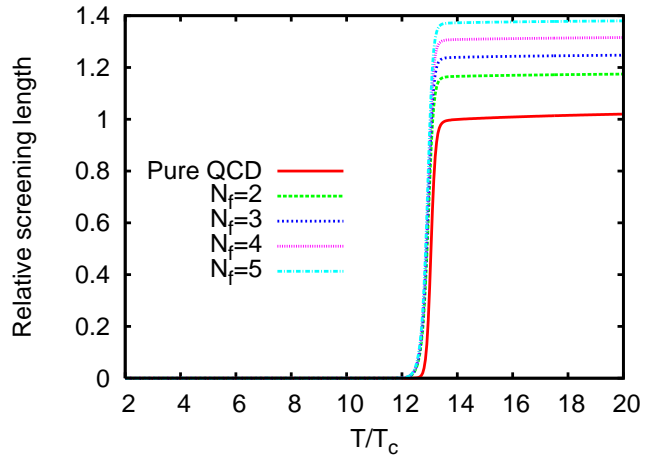}
\caption{(color online) The relative screening length ${\cal R}_{h/id}$ for  EOS2 as a function of temperature. }
\end{center}
\end{figure}
It appears that the perfect screening is indeed the strongest prediction of EOS2 and must be most easily tested in heavy ion collisions, where
temperatures upto $3T_c$ are expected at LHC. This is in sharp contrast with the assumptions of
a near ideal behaviour, and  also some theoretical analyses which in fact propose an enhanced
production of $J/\Psi$ at LHC energies \cite{lhc}. We, therefore, attempt to compare these predictions with the lattice results below.

\subsection{Comparison with the lattice results}
In this subsection, we compare our results on screening length of EOS1 and EOS2 with the   lattice results. 
Lattice compuattions extract the screening lengths from the quark-antiquark free energies.
To be concrete, we make the comparison with three distinct values of the coupling constant, $g'=.3, .5, .8$.

\begin{figure}[htb]
\begin{center}
\includegraphics[scale=.7]{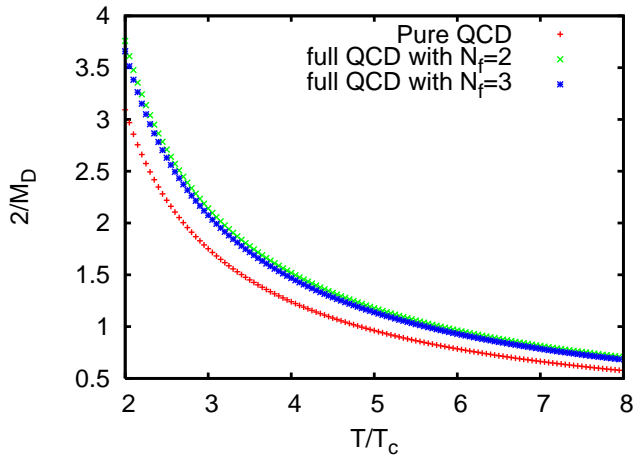}
\caption{(color online) Behaviour of $2/M_D$ with $T/T_c$ for $g^{\prime}=.3$ for EOS1. Note that $2/M_D$ is measured in {\it fm}}
\end{center}
\end{figure}

\begin{figure}[htb]
\begin{center}
\includegraphics[scale=.7]{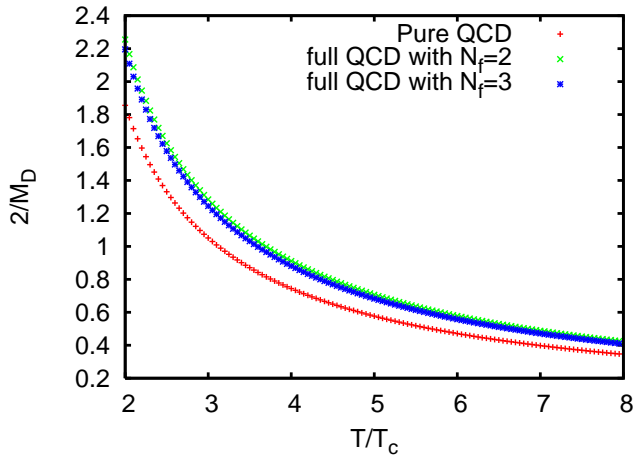}
\caption{(color online) Behaviour of $2/M_D$ with $T/T_c$ for $g^{\prime}=.5$ for EOS1. Note that $2/M_D$ is measured in {\it fm}}
\end{center}
\end{figure}
Further, we consider three cases, (i) pure QCD, (ii) $N_F=2$ and (iii) $N_F=3$. To facilitate a proper comparison, we take the respective transition temperatures to be $T_c = 270 MeV,~~ 203 MeV$
and  $195 MeV$, as given by lattice computations. The comparison is shown only with EOS1 since
EOS2 predicts absolute screening in the range $ 2T_c < T < 12T_c$ that we are interested in.
The results are shown in Figs. 13-15. As observed, the screening weakens with increasing $g^{\prime}$; moreover,
the screening weakens with the increase in the number of flavours as well. For an explicit comparison, we consider the results reported by Kaczmarek and  Zantow\cite{zantow} who detemine the screening length by identifying
it essentially with the first moment of the $q\bar{q}$ free energy. Their results are displayed in Fig. 2 of
\cite{zantow} to which we refer henceforth.
Interestingly, the same qualitative
features are exhibited EOS1 and the lattice results, in both the aspects, {\it viz}, the dependence on the
coupling constant as well as on the number of flavours.
However, the agreement fails to get quantitative. The lattice results  predict screening
lengths which are smaller in value, except for $N_F=3$ than the EOS1 results. Indeed, the lattice screening length
is $\sim 0.7 fm$ in the vicinity of $T_c$, and drops to $\sim 0.4 fm$ close to $2T_c$. It is evident from
Figs. 13-15 that the results of EOS1 are $3-10$ times higher in value. Any better agreement with a further increase in the value of $g^{\prime}$ is ruled out since$g^{\prime} \le 1$ necessarily.

\begin{figure}[htb]
\begin{center}
\includegraphics[scale=.7]{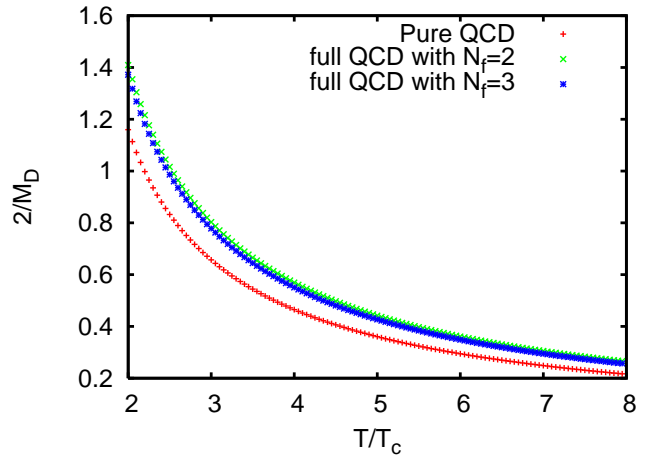}
\caption{(color online) Behaviour of $2/M_D$ with $T/T_c$ for $g^{\prime}=.8$ for EOS1. Note that $2/M_D$ is measured in {\it fm}}
\end{center}
\end{figure}

\section{ Conclusions and outlook}
 In conclusion,  we have extracted the distribution functions for gluons and quarks from two equations of state,  in terms of  effective chemical potentials for the partons. The chemical potentials
are shown to be highly sensitive to the inclusion of $O(g^6ln(1/g))$  contributions, and exhibited most vividly by the
screening length. Surprisingly, EOS2 which has interactions upto $O(g^6 ln(\frac{1}{g}))$ shows less nonideal behaviour compared to
EOS1 (which has contributions upto $O(g^5)$. Equally strikingly, the plasma corresponding to
EOS2 is predominantly gluonic, in the sense that the Debye mass from the gluonic sector diverges in the range $2T_c \le T \le 12T_c$. This result is in  contrast with the less
precise EOS1, where the gluonic contribution is not that overwhelming. 

To place our analysis in perspective, we note that our analysis is based on but two equations of state, neither of which has full non-perturbtive contributions. Nevertheless, it may not be without merit since EOS2, for instance, makes rather strong predictions which may be tested; EOS1 is, on the other hand,  seen to be in qualitative agreement   with the lattice results. Indeed,the work does provide a platform to study quantitatively the import of the EOS to heavy ion collisions in a quantitative manner. Experiments at 
 LHC may be able to probe these EOS since a temperature in the range  $T \sim 2-3T_c$ is expected to be
achieved there. More importantly, the method developed here can be easily employed to study more precise
EOS (as from lattice computations), or more general EOS (as the inclusion of baryonic chemical potentials). To be sure, an incisive analysis is possible only after studying other quantities such as 
the viscosity, its anomalous component \cite{arj,bm}, the viscosity to entropy ratio, and the specific heat. Finally,the insertion of the appropriate equilibrium distribution functions in the semiclassical transport equations  allow for studying (i) the production and the equilibration rates for the QGP in heavy ion collisions
\cite{vr1,vr2,vr3}, and (ii)  the color response functions\cite{akr}, of which the Debye mass is but one limiting parameter. These will be taken up in subsequent publications.
These investigations will be taken up separately.

\section{Acknowledgments}
We thank DDB Rao for assistance with numerical work.
Two of us, VC and RK, acknowledge CSIR (India) for financial support through the award of a fellowship.

\section{Appendix}
We use the following standard integrals while extracting effective chemical potential:
\begin{widetext}
\bea
\int_0^\infty p^2\frac{\exp(-p)}{(1-\exp(-p))^3} dp &=& 2(1+\sum_{n=1}^{\infty} (\frac{1}{(n+1)^3}\prod_{k=1}^n \frac{3+k-1}{k!}))\nn
\int_0^\infty p^2\frac{\exp(2p)}{(1+\exp(p))^3} dp &=& \frac{\pi^2}{12}+\log(2)\\
\int_0^\infty p^2\frac{\exp(p)}{(1-z\exp(p))^2} dp &=& \frac{PolyLog[2,z]}{z}\nn
\int_0^\infty p^2\frac{\exp(p)}{(1+z\exp(p))^2} dp &=& -\frac{PolyLog[2,-z]}{z}
\eea
\end{widetext}
The  coefficients in the perturbative expansion of $\log(Z_g)$ and $\log(Z_f)$ ,are as follows;
\bea
A^{(1)}_g &=& \frac{V}{2\pi^2 g_b} 2\zeta(3)\nn
A^{(1)}_f &=& \frac{V}{2\pi^2 g_f}\frac{3}{2}\zeta(3)\nn
A^{(2)}_g &=&\frac{V}{2\pi^2 g_b}\frac{\pi^3}{3}\nn
A^{(2)}_f &=& \frac{V}{2\pi^2 g_f}\frac{\pi^3}{6}\nn
A^{(3)}_f &=& \frac{V}{2\pi^2 g_f}2\log(2)\nn
A^{(3)}_g &=& \frac{V}{2\pi^2 g_b}[4(1+\sum_{n=1}^{\infty} (\frac{1}{(n+1)^3}\prod_{k=1}^n \frac{3+k-1}{k!}))-\frac{\pi^2}{3}]
\eea
where $g_b=8\times 2$ and $ g_f=6 N_f$ are the degeneracy factors for gluons and quarks.


\begin{thebibliography}{99}
\bibitem{comment1} The  expression for the QCD running coupling constant displayed in Eq.(\ref{eqn17}) is allowed in the region where the weak perturbative techniques are valid. Since, the weak coupling techniques give convergent results for more than $2T_c$, where $T_c$ is QCD transition temperature and a free parameter here. In this region, we can consider the asymptotic limit of the running coupling constant. Employing this, we shall see that the hot QCD EOS relative to ideal EOS and Effective Chemical Potentials scales with $T/T_c$.
\bibitem{comment2} The function $PolyLog[2,z]$  is defined by the series
 $PolyLog[2,z]=\sum_{k=1}^\infty \frac{z^k}{k^2}$, with a radius of convergence given by
 $\vert z \vert < 1$.
\end{thebibliography}
\end{document}